\documentclass[%
 reprint,
 amsmath,amssymb,
 aps,prl
]{revtex4-1}

\usepackage{color}
\usepackage{graphicx}
\usepackage{dcolumn}
\usepackage{bm}
\usepackage{wrapfig}

\begin{document}

\def\gpot{\phi}          
\def\dgpot{\psi}         
\def\cpot{\Phi}          
\def\dcpot{\Psi}         
\def\init{_I} 
\def\now{_F}  
\def\bx{{\bf x}}         
\def\bq{{\bf q}}         
\def\bv{{\bf v}}         
\def\bp{{\bf p}}         
\def\bg{{\bf g}}         
\def\everything{V}       

\def\mnras{MNRAS}

\def\be{\begin{equation}}
\def\ee{\end{equation}}
\def\ba{\begin{eqnarray}}
\def\ea{\end{eqnarray}}

\preprint{APS/123-QED}
\title{Accurate Baryon Acoustic Oscillations reconstruction via semi-discrete optimal transport}  
\author{Sebastian von Hausegger$^{1,2,3}$}
\email{sebastian.vonhausegger@physics.ox.ac.uk}
\author{Bruno Levy$^2$}
\author{Roya Mohayaee$^{1,3}$}
\affiliation{$^{1}$Rudolf Peierls Centre for Theoretical Physics, University of Oxford, Parks Road, Oxford OX1 3PU, United Kingdom}
\affiliation{$^{2}$Université de Lorraine, CNRS, Inria, LORIA, F-54000 Nancy, France}
\affiliation{$^{3}$Institut d'Astrophysique de Paris, CNRS, Sorbonne Université, 98bis Bld Arago, 75014 Paris, France}
\date{\today}

\begin{abstract}

Optimal transport theory has recently reemerged as a vastly resourceful field of mathematics with elegant applications across physics and computer science.  Harnessing methods from geometry processing, we report on the efficient implementation for a specific problem in cosmology --- the reconstruction of the linear density field from low redshifts, in particular the recovery of the Baryonic Acoustic Oscillation (BAO) scale.  We demonstrate our algorithm's accuracy by retrieving the BAO scale in noise-less cosmological simulations that are dedicated to cancel cosmic variance; we find uncertainties to be reduced by a factor of 4.3 compared with performing no reconstruction, and a factor of 3.1 compared with standard reconstruction.

\end{abstract}
                            
\maketitle

\section{Introduction}\label{sec:Introduction}

Linear perturbations in the primordial Universe propagate as sound waves through the photon-baryon plasma until light and matter decouple.  In a balance of radiation pressure and gravity, the baryon-to-photon and the matter-to-radiation ratios define a correlation length that is imprinted as a distinct feature onto the density fields, detectable in both the Cosmic Microwave Background (CMB) at early and the Large Scale Structure (LSS) at late times~\cite{1966JETP...22..241S,1970Ap&SS...7....3S,1970ApJ...162..815P}.  However, non-linear clustering of galaxies at low redshifts distort this imprint and therewith impedes unbiased detection~\cite{Eisenstein:2006nj}, calling for methods to undo these non-linear effects, to `reconstruct' the linear density field, and thereby to enable accurate and precise measurement~\cite{Eisenstein:2006nk}.  This so-called Baryon Acoustic Oscillation (BAO) scale serves as a unique tool to map the expansion history of the Universe, and hence has taken up a central role in cosmological analyses; most notably, its detection~\cite{2005MNRAS.362..505C,2005ApJ...633..560E} has provided constraints on standard quantities such as the Hubble constant and dark energy density~\cite{Zhao:2018gvb}, with promising capabilities to, with future observations, even constrain more nuanced theories of, e.g., light degrees-of-freedom~\cite{2017JCAP...11..007B}. Therefore, and especially in light of present and upcoming large galaxy surveys~\footnote{To date, these include, but are not limited to, \textit{e.g.}, DESI~\cite{2016arXiv161100036D}, LSST~\cite{2012arXiv1211.0310L}, Euclid~\cite{2011arXiv1110.3193L}.}, the development of fast, accurate, and scalable reconstruction methods is highly relevant to optimizing the yield of LSS studies. As a prime example of the wide and successful applicability of modern Optimal transport theory, this \textit{Letter} promotes a recent implementation of one such method.

Optimal transport theory describes mappings between probability measures that minimise a total cost function while satisfying a volume conservation constraint~\cite{OTON,opac-b1122739}.  Recent advances in both the mathematical and algorithmic aspects of the theory paved the way for breakthroughs in various fields, such as artificial intelligence, economics, meteorology, biology, and physics — due to the universality of processes minimizing an action.  In the context of this work, we exploit natural connections between optimal transport and physics~\cite{EURNature,2021MNRAS.506.1165L}: the to-be-minimised quantity and volume conservation are an action integral and the continuity equation respectively.  The Lagrange multiplier associated with the constraint turns out to be the gravitational potential.  Drawing from and building on all these advances, we recently developed a deterministic algorithm~\cite{2021MNRAS.506.1165L} that efficiently reconstructs the sought-for linear density field.  After having fully characterized its behavior in Ref.~\cite{2021MNRAS.506.1165L}, we here focus on the accuracy with which it retrieves the BAO scale by applying it to noise-free cosmic variance cancelling cosmological simulations generated with the \textsf{FastPM}~\cite{Feng:2016yqz} algorithm.

This \textit{Letter} recapitulates the mathematical basis of our method, illustrates its particular application to cosmological density field reconstruction, and finally showcases its excellent results in comparison with standard reconstruction by the use of noise-free cosmic variance cancelling cosmological simulations.

\section{Monge-Amp\`ere-Kantorovich Reconstruction}\label{sec:MAK_Reconstruction}

\begin{figure*}
\centerline{\includegraphics[width=\textwidth]{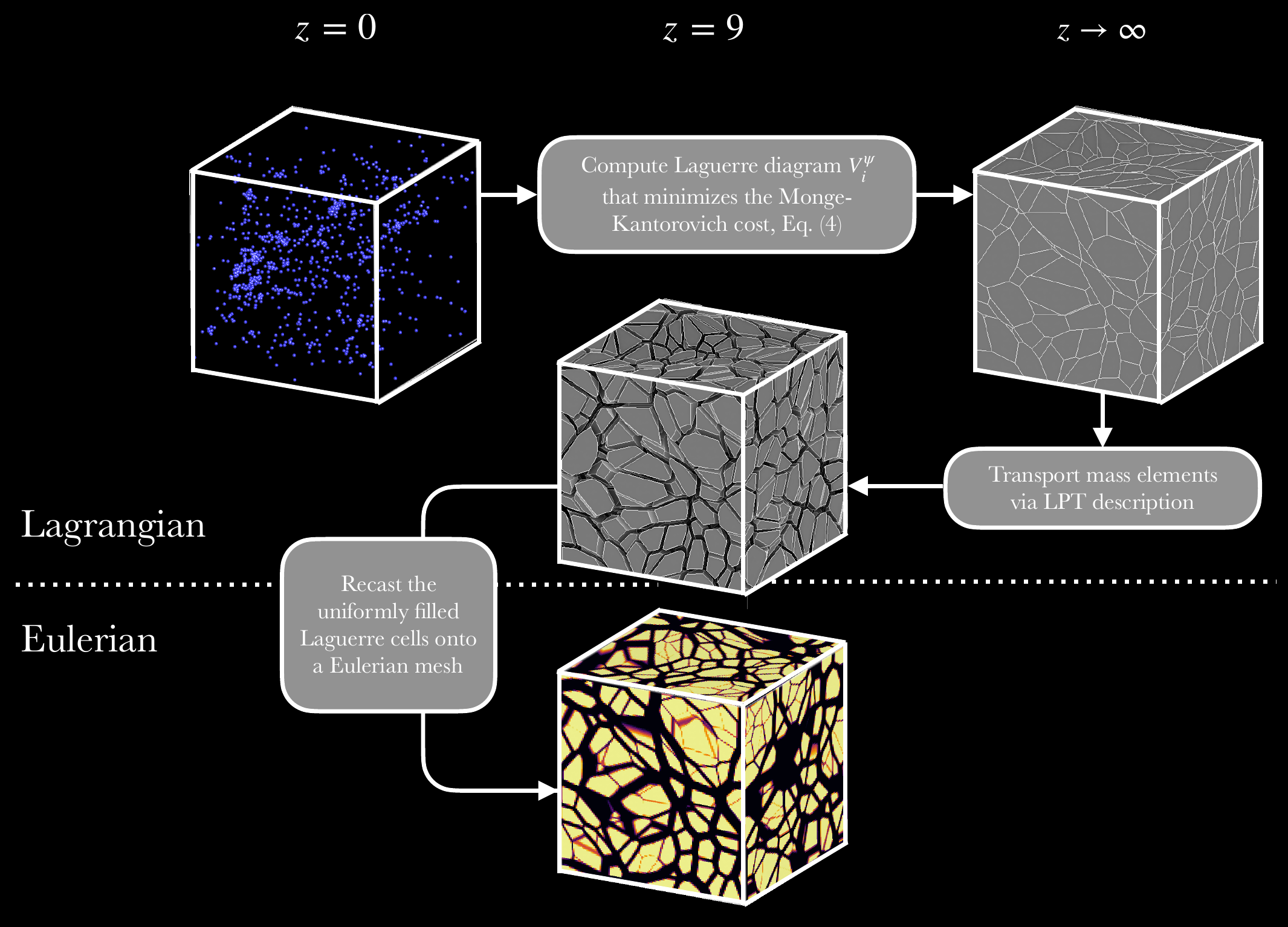}}
\caption{Semi-discrete Monge-Amp\'ere-Kantorovich Reconstruction.  Illustration of the reconstruction's flow beginning with an input of point masses at $\tau_F$ (here $z=0$) that are mapped to their corresponding regions at $\tau_I$ (here $z\rightarrow\infty$) from where they draw their mass through gravitational collapse.  The regions are the polyhedral cells defined by the Laguerre diagram as defined in the main body.  In a subsequent step the cells evolve until a time $\tau_F>\tau>\tau_I$ (here $z=9$) and according to a LPT description -- in the present paper the Zel'dovich approximation.  In the final step a mass density is expressed on a Eulerian mesh. 
The figure shows a tiny subset of a reconstruction from AbacusCosmos, that comprises 9.7 million Laguerre cells.
}
\label{fig:laguerrecellsonNbody}
\end{figure*}

Monge-Amp\`ere-Kantorovich (MAK) reconstruction uniquely determines the Lagrangian trajectories of a given Eulerian distribution of particle positions by solving an optimal transport (OT) problem~\cite{EUR,2021MNRAS.506.1165L}.  To this effect, consider self-gravitating matter in an Einstein-de~Sitter universe, where particle trajectories are the solution to extremizing the action $I$ subject to the Poisson equation, mass conservation and appropriate boundary conditions,
\begin{equation}
    I=\frac{1}{2}\int_{\tau_I}^{\tau_F}\int_\everything\,\left(\rho|\bv|^2 + \frac{3}{2}|\nabla_x\gpot|^2\right)\tau^{3/2}\;d^3x\,d\tau.
    \label{eq:fullaction}
\end{equation}
Here, $\bx=\bx(\bq,\tau)$ are the trajectories in Lagrangian coordinates of particles initially at $\bq$, $\bv$ is the Eulerian, co-moving velocity field, $\rho$ is the Eulerian density field, and $\gpot$ is the gravitational potential. Serving as a time variable, $\tau$ denotes the amplitude of the growing linear mode, normalized such that initially $\tau_I=0$ and finally $\tau_F=1$.  Correspondingly, the final density $\rho(\tau_F)$ is considered to have evolved from an initially uniform state, $\rho(\tau_I)=1$.  At final time $\tau_F$, we suppose that the density field is clustered into a set of points $({\bf x}_i)_{i=1}^N$ with masses $(m_i)_{i=1}^N$. 
 Ultimately, we aim to recover the (linear) density field at high redshift, or $\tau=\tau_I+\epsilon$, $\epsilon\ll1$, from a low-redshift, or present-time distribution of clustered matter.  In practice, this will require finding the optimal map that assigns initial positions $\bq$ to the input particle positions $\bx(\bq,\tau_F)$. 

Linearizing around the stationary points of the action, Eq.~(\ref{eq:fullaction}), one finds only the kinetic term to remain, resulting in uniform rectilinear motion~\cite{LaudauCourse}.  Interestingly, this case is proven~\cite{DBLP:journals/nm/BenamouB00} to be equivalent to the $L_2$ Monge-Kantorovich optimal transport problem~\cite{Monge1784},
\begin{equation}
    \inf_{\bx_F}\int_\everything\,\rho(\bq)|\bx_F(\bq)-\bq|^2\;d^3q,
    \label{eq:mongekantorovichcost}
\end{equation}
in which the integrated squared distance is minimized, subject to mass conservation, $\rho_F(\bx)/\rho_I(\bq)=\textrm{det}\left[d^3{\bf x}/d^3{\bf q}\right]$. In the dual formulation of the problem, that exchanges the unknown application $\bx_F$ with the mass conservation constraint, one solves the \emph{Monge-Ampère equation} for a scalar function $\Phi$, associated with the constraint, and called the \emph{Kantorovich potential}:
\begin{equation}
    \frac{\rho_F(\bx)}{\rho_I(\bq)}=\textrm{det}\left[\frac{\partial^2\Phi}{\partial {\bf q}_i\partial {\bf q}_j}\right],
\label{eqn:MA}    
\end{equation}
The optimal assignment map $\bq\mapsto\bx_F(\bq)$ turns out to be uniquely determined as the gradient of $\Phi$~\cite{BrenierPFMR91}.  
From the second-order optimality condition, one can prove that $\Phi$ is a \emph{convex} function, which is interesting because it implies the absence of shell-crossing during structure formation: intuitively, one may think of the velocity field as the normals to the graph of $\Phi$, and the normal vectors to a convex body do not intersect. More formally, the proof is obtained by contradiction: the existence of a collision would imply an inequality that would in turn contradict the convexity of $\Phi$ (details in~\cite{2021MNRAS.506.1165L}).

It is interesting to express the optimization problem as a function of physical quantities instead of $\Phi$, using the relation between $\Phi$, the initial gravitational potential $\gpot_I$ and the final gravitational potential $\gpot_F$:
\begin{align}
  \Phi(\bq) & = & 1/2\,\bq^2-\gpot_I(\bq) 
  \label{eqn:potpot} \\[2mm]
  \gpot_I(\bq) & = & \inf_{\bx} [1/2 |\bx - \bq|^2 - \gpot_F(\bx)].
  \label{eqn:Legendre}
\end{align}
The initial and final potentials $\gpot_I$ and $\gpot_F$ are mutually related through the \emph{Legendre-Fenchel transform} (Eq. \ref{eqn:Legendre}) (the same one that converts between Lagrangian and Hamiltonian mechanics). They turn out to be the solution of the following optimization problem:
\begin{equation}
    \max_{\gpot_I,\gpot_F} K = 
     \int_V \gpot_I({\bf q}) \rho_I({\bf q}) d{\bf q} + 
     \int_V \gpot_F({\bf x})\rho_F({\bf x})d{\bf x},
     \label{eqn:K}
\end{equation}
where the functional $K$ is called the \emph{Kantorovich dual}.
 

While numerical methods for approximately solving the Monge-Amp\`ere (equation \ref{eqn:MA}) have been devised~\cite{2018PhRvD..97b3505S,2019MNRAS.483.5267B,2021ApJS..254....4L}, elegant algorithmic developments in computer science allow for efficiently constructing exact solutions (see \cite{ComputationalOT} for a review).

Previous reconstruction methods that exactly solve the Monge-Amp\`ere equation \cite{EUR,EURNature} considered a \emph{discrete} version of the Monge-Kantorovich problem (Eq. \ref{eq:mongekantorovichcost}), i.e.~to find the permutation $j(i)$ between a finite set of $N$ homogeneously distributed particle positions $(\bq_j)_{j=1}^N$ at $\tau_I$ and their corresponding positions $(\bx_i)_{i=1}^N$ at $\tau_F$ that minimizes $\sum_i |\bx_{i(j)} - \bq_j |^2$.  Efficient combinatorial methods \cite{Bertsekas1992} avoid exploring the full set of $N!$ possible permutations; however, they still scale as $\sim\!N^2\log(N)$, rendering such algorithms increasingly unfeasible for larger data sets. Moreover, using these combinatorial methods, it is difficult to allocate a different mass $m_i$ to each point.

However, exploiting the variational nature of the Monge-Kantorovich problem, the present \emph{semi-discrete} approach replaces this exhaustive combinatorial search by the optimization of a well-behaved objective function, simultaneously making use of the geometric structure of the setting: Instead of representing the initial condition as a discrete set of points $(\bq_j)_{j=1}^N$, we consider it as a continuum, while the density at $\tau_F$ is concentrated on a set of points $\bx_i$ with the associated masses $m_i$. Replacing $\gpot_I$ by its expression (Legendre transform of $\gpot_F$, Eq. \ref{eqn:Legendre}), the Kantorovich dual $K$ in Equation \ref{eqn:K} becomes:
\begin{equation}
    K(\psi) = \sum_i \int_{\everything^\psi_i} \left[\frac{1}{2}|\bx_i - \bq|^2 - \psi_i\right]\; d^3q + \sum_i m_i \psi_i,
\label{eqn:SDK}
\end{equation}
that depends on the vector $\psi = (\gpot_F(\bx_i))_1^N$, and where the subsets $V_i^\psi$, also stemming from the Legendre-Fenchel transform of $\gpot_F$, are defined by:
\begin{equation}
    \everything^\psi_i = \left\{ \bq \ \left| \ \frac{1}{2}|\bx_i - \bq|^2 - \psi_i < \frac{1}{2}|\bx_j - \bq|^2 - \psi_j, \ \forall j \neq i\right. \right\}.
    \label{eq:laguerrecell}
\end{equation}
The regions $V_i^{\psi}$ are called \emph{Laguerre cells}, and they form a \emph{Laguerre diagram} (more on this in \cite{DBLP:conf/compgeom/AurenhammerHA92,DBLP:journals/cgf/Merigot11,journals/M2AN/LevyNAL15,OTReviewMerigotThibert}). Each Laguerre cell $V_i^\psi$ corresponds to the region of space mapped to $\bx_i$ through the reconstructed motion.

The vector $\psi$ that determines the Laguerre diagram can be obtained by maximizing the Kantorovich dual $K$ (Eq. \ref{eqn:SDK})  
subject to the constraint that no cell is empty ($V^\psi_i \neq \emptyset, \; \forall i$).  Finally, one retrieves the gravitational potential at $\tau_I$ through the Legendre
transform of $\psi$: 
\begin{equation}
\begin{aligned}
    \gpot(\tau_I,\bq)&=\inf_{\bx_F}\left[\frac{1}{2}\left|\bq-\bx_i\right|^2-\psi_i\right]\\
    &=\frac{1}{2}\left|\bq-\bx_{i(\bq)}\right|^2-\psi_{i(\bq)},
\end{aligned}
\end{equation}
where $i(\bq)$ is the index of the Laguerre cell $V_i$ that contains $\bq$.  In other words, maximizing $K(\psi)$ is equivalent to solving for the Kantorovich potential $\Phi$ in the Monge-Amp\`ere equation.

The Kantorovich dual $K$ has two properties that are important from the point of view of numerical optimization. Firstly, $K$ corresponds to the lower envelope of a family of affine functions, hence it is \emph{concave}~\cite{DBLP:conf/compgeom/AurenhammerHA92,DBLP:journals/cgf/Merigot11,journals/M2AN/LevyNAL15,OTReviewMerigotThibert}, which ensures
the existence and uniqueness of $\psi$. Secondly, by studying the Taylor expansion of $K$ in two configurations differing from a combinatorial change in the Laguerre diagram, one can observe that the terms up to $2^{nd}$ order match~\cite{DBLP:journals/tog/LiuWLSYLY09}, hence $K$ 
is $C^2$\emph{-smooth}, which allows for an efficient and convergent optimization~\cite{DBLP:journals/corr/KitagawaMT16}.  It is the combination of the two analytic properties of $K$ (convexity and smoothness) that allow replacing the expensive combinatorial computation of previous methods with a Newton method that exploits the first- and second-order derivatives of $K$ to efficiently minimize it.

Finally, each Laguerre cell $V_i^\psi$ at $\tau_I$ is mapped to its corresponding point $\bx_i$ at $\tau_F$.  At an intermediate time $\tau$ the trajectories of the mass elements can be deduced via a Lagrangian perturbation theory (LPT) description, in our case the Zel'dovich approximation~\cite{zeldo:1970}.  This, in turn, can be converted as a mass density on a Eulerian grid, as depicted in Fig.~\ref{fig:laguerrecellsonNbody} (see also~\cite{2021MNRAS.506.1165L}).

\begin{figure*}
\centerline{\includegraphics[width=\textwidth]{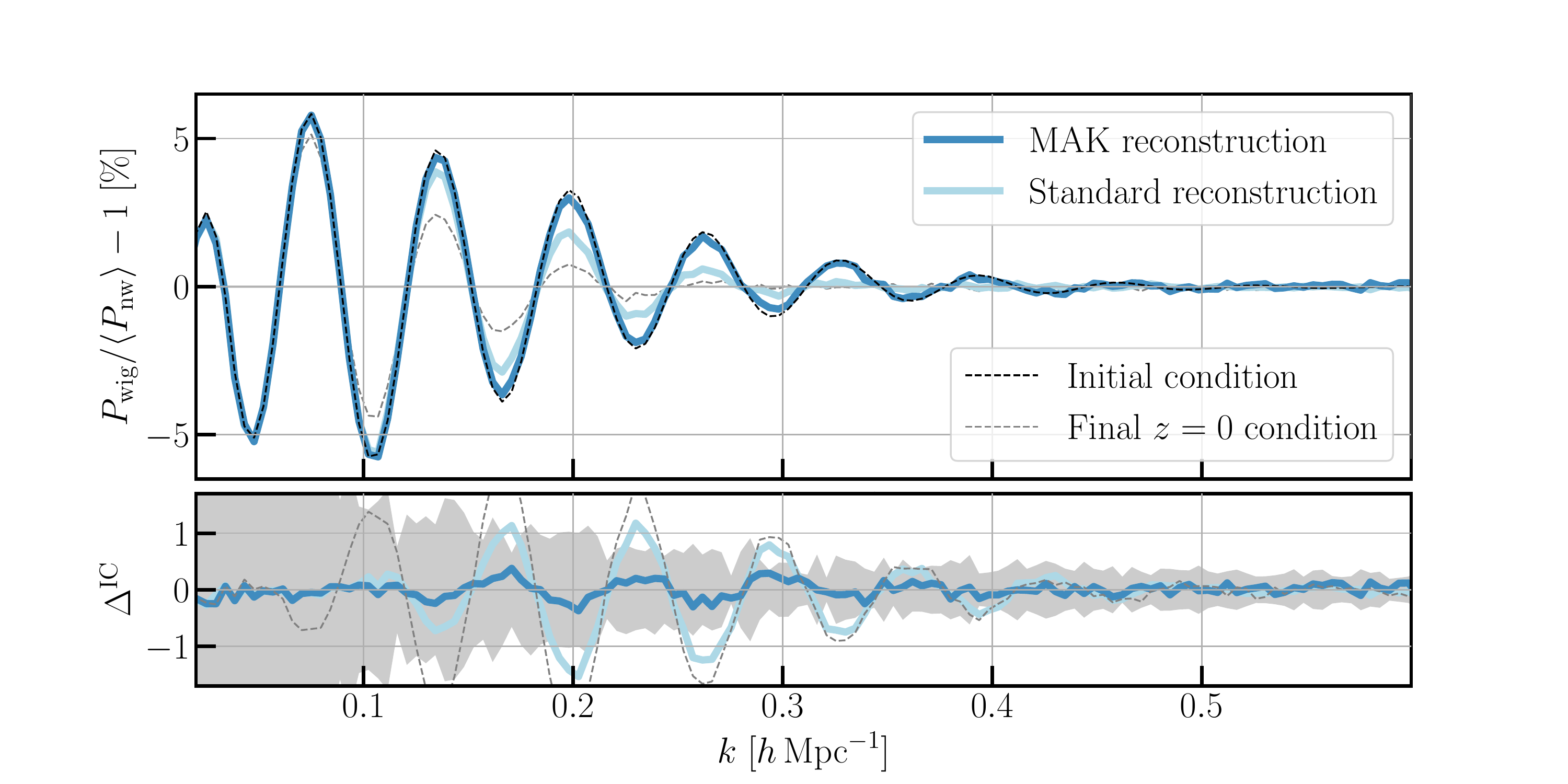}}
\caption{Fractional BAO signal in simulations and reconstructions.  \textit{Top panel:} Ratio of power spectra from simulations with and without BAO signal, averaged over ten \textsf{FastPM} simulations and their reconstructions, $\langle P_{\rm wig}(k)-P_{\rm nw}(k)\rangle/\langle P_{\rm wig}(k)\rangle$.  \textit{Bottom panel:} Deviations from the initial condition's fractional BAO signal, $\Delta^{\rm IC}(k) = \langle P_{\rm wig}(k)\rangle / \langle P_{\rm nw}(k)\rangle-\langle P_{\rm wig}^{\rm IC}(k)\rangle / \langle P_{\rm nw}^{\rm IC}(k)\rangle$.  The shaded band indicates the standard deviation $\sigma^{\rm IC}(k)$ as estimated from the simulations and defined in the text.}
\label{fig:PowerComparison}
\end{figure*}

\section{BAO reconstruction}\label{sec:BAO_Reconstruction}

Baryonic Acoustic Oscillations imprint a signature in the matter power spectrum by periodically modulating the large-scale power at a frequency corresponding to the sound horizon at decoupling~\cite{1966JETP...22..241S}.  Non-linear gravitational evolution blurs and shifts this feature~\cite{Eisenstein:2006nj}, complicating reliable detection and interpretation, wherefore reconstruction algorithms were devised that correct for such influences.  In addition to physical effects, finite survey volumes, both in simulations and in practice, induce sample variance (so-called cosmic variance) into the measured quantities due to a reduced number of modes on these very scales.  Since these modulations can be seen as only small perturbations to the  bulk of the power spectrum, the variance is dominated by the un-modulated power spectrum, suites of dedicated $N$-body simulations can effectively cancel cosmic variance for analyses that only target the BAO signal~\cite{Prada:2014bra,2017PhRvD..96b3505S,Wang:2017jeq}, e.g., to test reconstruction methods.  In this view, we apply our reconstruction algorithm~\cite{2021MNRAS.506.1165L} to the \textsf{FastPM}~\cite{Feng:2016yqz} simulations of Ref.~\cite{Ding:2017gad}.

This suite of \textsf{FastPM} simulations is comprised of pairs of simulations initiated with the same random phases at redshift $z=9$, yet with power spectra that differ by the presence of the BAO feature, in the following referred to as `wiggle' and `no-wiggle' power spectra, or $P_{\rm wig}(k)$ and $P_{\rm nw}(k)$.  Each of the simulations traces $2048^3$ particles in a cube of $1380\,h^{-1}$Mpc side length from $z=9$ to $z=0$, and saved at redshift $z=0$ as a particle sample.  In order to keep computing time low, we sub-sample only $\sim\!1\%$ of all particles in each of the ten simulation pairs we consider, resulting in about $85\times10^6$ particles per simulation.  All simulations follow a $\Lambda$CDM cosmology with parameters from Ref.~\cite{Planck:2015fie}, which gives an expected BAO scale of $r_{\rm BAO}^{\rm th}=147.5\,\rm{Mpc}$.

Beginning with these particles' positions, $\bx$, at $z=0$, we compute the density field~\cite{2021MNRAS.506.1165L} at $z=9$ as described in the previous section and via the choice of an appropriate linear growth factor~\footnote{The present analysis is not very sensitive to the choice of growth factor~\cite{Lukic:2007fc} as it is only concerned with the ratio of `wiggle' and `no-wiggle' simulations.  Nevertheless, when dealing with real data, a slight mismatch between the true growth factor and a perhaps wrongly approximated one will in effect show up as a bias factor, $b$, to the reconstructed power, and can therefore easily be taken into account in a subsequent step, e.g. in a parameter fit one can marginalize over this parameter.}.  Figure~\ref{fig:PowerComparison} demonstrates the effectiveness of our reconstruction, where the relative differences, $d(k)=P_{\rm wig}(k)/\langle P_{\rm nw}(k)\rangle-1$, of initial ($z=9$) and final ($z=0$) power spectra are compared with those of our reconstructed density fields, averaged over all ten simulations.  Our reconstruction's excellent performance is further highlighted by contrasting it against the result obtained from so-called standard reconstruction~\cite{Eisenstein:2006nk}\footnote{We computed the standard reconstructed density field using the routine \texttt{FFTRecon} of the \texttt{python} suite \texttt{nbodykit}~\cite{Hand:2017pqn}.}, visible through the agreement of initial condition and reconstruction out to wave numbers $k\gg0.15\,h^{-1}\textrm{Mpc}$, where discrepancies between initial condition and standard reconstruction first appear.

\begin{table*}
\centering
\caption{Bias and uncertainties of BAO scales recovered in the fractional power spectra of simulations and reconstructions with and without cancelling cosmic variance.  The columns lists the mean values and standard deviations of $r_{\rm BAO}$ (\textit{left columns}) $r_{\rm BAO}-r_{\rm BAO}^{\rm IC}$ (\textit{right columns}) obtained from fitting $\alpha$ to the power spectra as described in the main body.}
\begin{tabular}{lllll}
&   \multicolumn{2}{c}{$\qquad\langle\hat r_{\rm BAO}\rangle$ $[\textrm{Mpc}]$} &   \multicolumn{2}{c}{$\qquad\sigma_{\hat r_{\rm BAO}}$ $[\textrm{Mpc}]$}    \\
&\qquad vs.~theory&\qquad vs.~IC&\qquad vs.~theory&\qquad vs.~IC\\
\hline
Initial cond.	&\qquad	$+0.02$	[$+0.01\%$]	&\qquad	$\pm0.00$	[$\pm0.00\%$]	&\qquad	$0.34$	[$0.23\%$]	&\qquad	$0.00$	[$0.00\%$]	\\
\hline
MAK rec.	&\qquad	$-0.11$	[$-0.08\%$]	&\qquad	$-0.13$	[$-0.09\%$]	&\qquad	$0.44$	[$0.30\%$]	&\qquad	$0.29$	[$0.19\%$]	\\
Standard rec.	&\qquad	$-0.03$	[$-0.02\%$]	&\qquad	$-0.05$	[$-0.03\%$]	&\qquad	$0.69$	[$0.47\%$]	&\qquad	$0.85$	[$0.58\%$]	\\
Final cond.	&\qquad	$+0.48$	[$+0.33\%$]	&\qquad	$+0.46$	[$+0.31\%$]	&\qquad	$1.06$	[$0.72\%$]	&\qquad	$1.19$	[$0.81\%$]
\end{tabular}
\label{tab:Results}
\end{table*}

In line with Ref.~\cite{2017PhRvD..96b3505S} we $\chi^2$-fit templates, $m(k)$, to the power spectrum ratios to obtain estimates of the BAO scale in each of the ten simulations.  We define $m(k)=e^{k^2\Sigma^2/2}\left[P^{\rm IC}_{\rm wig}(k/\alpha)/P^{\rm IC}_{\rm nw}(k/\alpha)-1\right]$ as the relative difference of power spectra of the initial, linear density field, allowing for a shift $\alpha=\hat r_{\rm BAO}/r_{\rm BAO}^{\rm th}$ of the BAO scale, and a Gaussian damping $\Sigma$~\footnote{We define $P_{\rm wig}^{\rm IC}(k/\alpha)/P_{\rm nw}^{\rm IC}(k/\alpha)$ as the average over all ten simulations, shown in Fig.~\ref{fig:PowerComparison}, and interpolated to accommodate values of $\alpha\neq1$.}, and up-weigh small scales via choice of the standard error $\sigma^2(k)=2\,\textrm{Var}\left[d(k)\right]/N_{\rm modes}(k)$, where $N_{\rm modes}(k)$ is the number of Fourier modes that contributes to the computed power in each $k$-bin.  We perform the fits over the full $k$-range shown in Figure~\ref{fig:PowerComparison}, and the best-fit values of $\alpha$ define the best-fit BAO scales $\hat r_{\rm BAO}$ in each fractional power spectrum $d(k)$.  Table~\ref{tab:Results} presents biases and uncertainties in retrieving $r^{\rm th}_{\rm BAO}$ in each of the simulations and reconstructions.  While comparison with the theory value $r_{\rm BAO}^{\rm th}$ (left columns) confirms and restates more precisely the results of Ref.~\cite{2021MNRAS.506.1165L}, the right columns optimally and for the first time showcase our algorithm's accuracy and precision in reconstructing the BAO scale from noiseless cosmological simulations, by subtracting from each simulation the inherent BAO scale,  $r_{\rm BAO}^{\rm IC}$, before determining mean and spread, thereby cancelling cosmic variance.

Due to the arising of shift terms in the non-linear power spectra~\cite{Eisenstein:2006nj,2012PhRvD..85j3523S,2017PhRvD..96b3505S}, the BAO scale at $z=0$ appears biased by $\sim\!0.3\%$, in accordance with previous findings~\cite{2010ApJ...720.1650S,2011ApJ...734...94M,2017PhRvD..96b3505S}.  This is accompanied by a $\sim\!0.8\%$ uncertainty that reflects the blurring of the BAO peak that as well is caused by non-linear gravitational growth.  Reconstruction reduces this bias~\cite{2012PhRvD..85j3523S} as we too see in both MAK and standard reconstruction, and further sharpens the BAO peak increasing the precision with which $\hat r_{\rm BAO}$ is determined; compared with the inherent uncertainty the simulations carry at the final condition -- including cosmic variance -- standard reconstruction improves the precision by a factor of $1.5$ while MAK reconstruction gives a factor of $2.4$ of enhancement.  In an idealised scenario, without cosmic variance, the factor $1.4$ improvement of standard reconstruction is surpassed by MAK reconstruction by as much as $4.3$.  In all cases we find a significant reduction of the bias as well.

As elaborated in Ref.~\cite{2021MNRAS.506.1165L}, subsample variance 
has significant impact on the overall error budget.  Both shot noise and subsample variance are virtually removed by the use of the present simulation suite, and cosmic variance is further cancelled by direct simulation-to-simulation comparison as we display in the right columns, Table~\ref{tab:Results}.  We therewith optimally test our reconstructions' accuracy. 

\section{Conclusions}\label{sec:Conclusions}

This \textit{Letter} demonstrates the application of Optimal Transport theory to a specific problem in cosmology, the reconstruction of the BAO peak in the matter power spectrum from low-$z$ observations.  BAO analyses play a crucial role in inferring cosmological parameters, and reconstruction methods have long aided the accuracy with which this signal is extracted.  Outperforming many of the most promising algorithms, our method scales well ($\propto N\log N$) with increased survey size, securing bright prospects in light of upcoming large-scale galaxy surveys.

In specific, we found that our reconstruction improves on detecting the BAO signal in the power spectrum by a factor of 4.3 compared with attempting to extract the BAO scale without having performed any reconstruction.  Even in the case of having applied the so-called standard reconstruction technique, our method reduces the uncertainties by more than a factor of 3.  This is highly promising especially given that in moving forward in time, we considered no more than the Zel'dovich approximation.  We therefore highly anticipate further improvement of reported accuracy by amending the second step in Fig.~\ref{fig:laguerrecellsonNbody} with corrections from higher-order Lagrangian perturbation theory~\cite{Bernardeau:2001qr,2017PhRvD..96b3505S}.

The next steps for optimal incorporation of our reconstruction method into analyses of survey data include its adaptation to account for the surveys' selection functions, halo masses, redshift-space distortions, and characterization and computation of the reconstruction covariance matrices.  Our algorithm's flexibility easily accommodates such modifications without losing its efficiency.

In summary, our method makes direct use of the variational nature of gravitational evolution and thereby its reconstruction. It finds a quick path to the solution by leveraging first and second order information of the problem (\textit{i.e.}~it being both smooth ($C^2$) and convex), while existing MAK methods need to exhaustively explore a huge $(N^2\log(N))$ combinatorial space. This is made possible by a fortuitous yet elegant convergence between the physical, mathematical and computational aspects of the problem: the specific cosmological setting that we considered (continuous mass transported to a point set) has nice mathematical properties (semi-discrete Monge-Amp\`ere equation translated into a smooth and concave optimization problem), with an underlying geometric structure (Laguerre diagram) that can be \emph{exactly} computed by our algorithm.

\begin{acknowledgments}
The authors thank Zhejie Ding for sharing their simulations. SvH thanks Yu Feng for assistance with FastPM.  BL and SvH acknowledge support from an Inria internal grant (Action exploratoire AeX EXPLORAGRAM).  SvH is supported at Oxford by the Carlsberg Foundation, and wishes to thank Linacre College for the award of a Junior Research Fellowship.  RM thanks the Rudolf Peierls Centre for hospitality.
\end{acknowledgments}

\bibliography{BAO}

\end{document}